# Simulating the future urban growth in Xiongan New Area: a upcoming big city in China

Xun Liang

liangx27@mail2.sysu.edu.cn

## Abstract

China made the announcement to create the Xiongan New Area in Hebei in April 1, 2017. Thus a new megacity about 110km southwest of Beijing will emerge. Xiongan New Area is of great practical significance and historical significance for transferring Beijing's non-capital function. Simulating the urban dynamics in Xiongan New Area can help planners to decide where to build the new urban and further manage the future urban growth. However, only a little researches focus on the future urban development in Xiongan New Area. In addition, previous models are unable to simulate the urban dynamics in Xiongan New Area. Because there are no original high density urban for these models to learn the transition rules. In this study, we proposed a C-FLUS model to solve such problems. This framework was implemented by coupling a fuzzy C-mean algorithm and a modified Cellular automata (CA). An elaborately designed random planted seeds mechanism based on local maximums is addressed in the CA model to better simulate the occurrence of the new urban. Through an analysis of the current driving forces, the C-FLUS can detect the potential start zone and simulate the urban development under different scenarios in Xiongan New Area. Our study shows that the new urban is most likely to occur in northwest of Xiongxian, and it will rapidly extend to Rongcheng and Anxin until almost cover the northern part of Xiongan New Area. Moreover, the method can help planners to evaluate the impact of urban expansion in Xiongan New Area.

# 1. Introduction

Chinese government decided to establish a state-level new development areas in Xiongan New Area in April 1, 2017, which is located in Hebei province in China. This area mainly includes 3 counties: Xiongxian, Rongcheng, Anxin and their surroundings (Jiang et al., 2017). It means that a new megacity like Shenzhen Special Zone and Pudong New Area in Shanghai is most likely to emerge in the future. As part of measures to advance the coordinated development of the Beijing-Tianjin-Hebei region, Xiongan is about to construct a triangular structure urban agglomeration with Beijing and Tianjin, which is similar to Preal River Delta urban agglomeration and Yangtze River Delta urban agglomeration, the most two economically advanced regions in China. According to the circular, the Xiongan New Area will cover around 100 square km initially and will be expanded to 200 square km in the mid-term and about 2,000 square km in the long term (Jiang et al., 2017; Yi-Qing and Liu, 2017).

One of the important functions of Xiongan New Area is to evacuate some non-capital functions from Beijing such as high energy consumption industry, enterprises with less innovation and technology, logistics base, back-up services of financial company and etc. Through this way, Chinese government hopes to explore a new manner of optimized development in high populated areas, and restructure the urban layout in the Beijing-Tianjin-Hebei region. Thus, the establishment of Xiongan New Area is of great significant both in reality and in history (Yi-Qing and Liu, 2017). However, researches about the urban development in Xiongan New Area are very limited.

Therefore, it is very important for national planners and decision makes to prevent poor urban designs (Clarke, 2014) for Xiongan New Area in the early stage of policy making. To reach that goal, basic and questions, such as 'if the new urban area should be established based on the original urban land?'; if not, 'where to build the new urban land in Xiongan New Area at the starting stage?'; and 'how the city in Xiongan New Area will develop?'; should be well answered. Similar questions about urban development in other regions can be analyzed by finding out the most potential development area, and furthermore projecting the possible urban development trajectory in Xiongan New Area by using Cellular Automata (CA) models (Liu et al., 2017), which can provide better understanding of the future urban dynamics and assist urban planners to test, explore and evaluate the outcomes of different planning policies (Chen et al., 2016).

As a spatially explicit models, CA models show advantages in urban modeling because it have self-organization capabilities and support spatial interactions (Clarke and Gaydos, 1998; Li et al., 2014) . CA models are attractive for urban change modeling as they are intrinsically spatial and easy to implement. In addition, macro policy (Liu et al., 2017) and spatial constraints (Yang et al., 2006) were also imposed by researchers to make CA better for practical uses (Chen et al., 2014). It have been widely used for simulating urban phenomena and help planners in making more reasonable land-use regulations and environmental protection policies (Yao et al., 2017) .

Over the past two decades, various kinds of CA models have been developed for simulating urban growth processes (Bren D Amour et al., 2017; Deal and Schunk, 2004; Jokar Arsanjani et al., 2018). Among these studies, the CA models based on the analysis of the historic urban form and various driving factors are interesting and useful for providing the insights of the nature of urban land-use dynamics (Guy et al., 1997; Li and Yeh, 2002;Verburg et al., 2002; Liu, 2006). They generally obtain transition rules or transition potential by using intelligence algorithm. These empirical transition potential models include logistic-regression (Chen et al., 2013), support vector machine (Ke et al., 2017), random forest (Kamusoko and Gamba, 2015b), artificial neural network (Li and Yeh, 2001) and etc. These kind of CA models can help researches obtain robust explanations of land-use pattern (Sohl et al., 2007) and are easy to be used (Verburg et al., 2004).

Although widely used in urban dynamics simulation, previous CA model based on transition potential models can only simulate the urban change in the region that have already developed for a period of history, because future urbanization process of these regions will most probability to develop around the developed high-density urban land or large urban clusters (Fragkias and Seto, 2009). And these CA models rely on enough historic land use data of study region to learn the urban transition potential or the probability-of-occurrence (Li et al., 2017). However, present Xiongan New Area is constructed by three developing counties with only 32 square km low-density urban land in 2013, and there are no predominant urban cluster in this region (Jiang et al.,

2017). Thus previous research framework that learns transition rules by mining the relationship between low-dense urban pattern and various driving factors with empirical supervisor method is not suitable for Xiongan New Area.

Therefore, the purpose of this paper is to propose an approach for simulating urban growth in Xiongan New Area without training transition rules from historic land use. Through the analysis of the various driving factor with a fuzzy C-mean (FCM) clustering algorithm, this method can identify the most potential hot-pot area and obtain urban transition potential without taking present low-dense urban land into account. Based on the urban transition potential, a modified version of CA-based FLUS (Future Land Use Simulation) model based on fuzzy C-mean clustering algorithm (C-FLUS) is employed to project the future urban development in study region. In the remainder of this paper, we describe our method, followed by our results and discussions.

## 2. Methods

**2.1. The fuzzy C-mean algorithm**

Fuzzy C-mean algorithm is one of the best known soft clustering algorithm (Rahimi et al., 2004). Different from the most commonly used K-mean method or ISODATA clustering that directly assign specified category for each sample, the fuzzy C-mean allows intermediate values between different classifications, it simultaneously gives the memberships of different categories for the same sample. The FCM algorithm can be expressed as the minimization of the following objective function $J_m$:

$$J_m = \sum_{i=1}^{N} \sum_{j=1}^{C} \mu_{ij}{}^m ||x_i - v_j||^2 \tag{1}$$

Where $x_i$ is one of the data points that with several dimensions or features; $v_j$ denotes the centroid of class j; $\mu_{ij}$ is a membership function that represents the degree of membership of $x_i$ in the cluster $j$, such that $\sum_{j=1}^{C} \mu_{ij} = 1$; $m$ is any real number greater than 1 which is an additional weighed exponent for the fuzzy membership. The operator $||\cdot||$ can be any type of inner product norm, typically the Euclidean norm is used (Li et al., 2003).

The membership function $\mu_{ij}$ defines the fuzziness of an image and also to define the information contained in the image (Selvakumar et al., 2012), which can be given by:

$$\mu_{ij} = \frac{1}{\sum_{j=1}^{C} \left( \frac{||x_i - v_l||}{||x_i - v_j||} \right)^{\frac{1}{m-1}}} \tag{2}$$

Where $1 \leq l \leq C$, $1 \leq i \leq N$. C is the number of categories and N is the amount of all the cluster samples.

The update of centers $v_l$ based on the following function:

$$v_l = \frac{\sum_{i=1}^{N} x_i \cdot \mu_{il}^m}{\sum_{i=1}^{N} \mu_{il}^m} \tag{3}$$

The function J is minimized by using the following equation for updating the membership degrees and the clustering centers iteratively until:

$$Max_{ij} = \{|\mu_{ij}^{(k+1)} - \mu_{ij}^{(k)}|\} < \varepsilon \tag{4}$$

Where $\varepsilon$ denotes the termination value or constant between 0 and 1, k is the number of iteration steps. The fuzzy C-mean algorithm products the membership of urban land based on various driving factors, which is regarded as the transition potential of future urban area in this study.

## 2.2. The C-FLUS model

Based on the transition potential achieved by FCM, the FLUS model is modified and implemented in the spatial simulation process. The FLUS model adopted in this study was designed by Liu et al. (2017). It is a CA-based method that is able to simulate complex land use change at different scale (Li et al., 2017). The original version of FLUS model is employed by training a neural network to achieve an urban probability-of-occurrence surface, then an improved CA model is implemented in spatial simulation process. In this study, the C-FLUS model replaces the neural network with the FCM algorithm for learning the relationships from and the various driving factors that drive the urban growth without considering the historic land use data.

### 2.2.1. Spatial simulation module

In the CA simulation process, the C-FLUS model employ a self-adaptive inertia coefficient to modify the total probability of urban land and non-urban according to their future amounts designed by decision makers. This coefficient can be given by:

$$\text{Inertia}_k^t = \begin{cases} \text{Inertia}_k^{t-1} & if\ |D_k^{t-1}| \leq |D_k^{t-2}| \\ \text{Inertia}_k^{t-1} \times \frac{D_k^{t-2}}{D_k^{t-1}} & if\ D_k^{t-1} < D_k^{t-2} < 0 \\ \text{Inertia}_k^{t-1} \times \frac{D_k^{t-1}}{D_k^{t-2}} & if\ 0 < D_k^{t-2} < D_k^{t-1} \end{cases} \quad (5)$$

where k is a value of 2, which means that only two land uses (urban land and non-urban land) are taken into account in the simulation. $Inertia_k^t$ denotes the inertia coefficient for land use type k at iteration time t. The $D_k^{t-1}$ is the difference between land use demand and allocated area at time t-1. Since the inertia coefficient is only defined for the land use type occupying the grid cell, if the potential land use type k is not the same as the current land use type c, the inertia coefficient of land use k will be defined as 1 and have no effect on total probability of land use type k for this grid cell.

Then the combined probability of all land-use types at each specific grid cell can be estimated with the following formula:

$$TP_{i,k}^t = P_{i,k} \times \Omega_{i,k}^t \times inertia_k^t \times con_{c \to k} \tag{6}$$

where $TP_{i,k}^t$ is the combined probability of the grid cell i to covert from the original land use into the target one k at iteration time t (only non-urban can convert to urban land in this study); $P_{i,k}$ denotes the transition potential of land use type k on grid cell i that was generated by FCM algorithm; $con_{c \to k}$ is a transition matrix that defines the possibility of conversion from the original land use type c to the target one k (1 denotes possible conversion and 0 denotes impossible conversion), and $\Omega_{i,k}^t$ denotes the neighborhood effect (percentage) of land use type k on grid cell i at time t, the neighborhood development density for land use type k is defined as:

$$\Omega_{p,k}^t = \frac{\sum_{N \times N} con(c_p^{t-1} = k)}{N \times N - 1} \quad N \geq 3, \ N \ mod \ 2 \neq 0 \tag{7}$$

In this equation, $\sum_{N \times N} con(c_p^{t-1} = k)$ represents the total number of grid cells occupied by the land use type *k* at the last iteration time *t - 1* within the $N \times N$ window. In the following, the urban land and non-urban compete with each other through a roulette constructed by the two combined probability in each cell. The roulette selection mechanism enables the C-FLUS model to better simulate the uncertainty and randomness in the urban growth process (Chen et al., 2013).

**2.2.2. Random planted seeds based on local maximums**

To better simulate urban development in the region that has no predominant urban and is specified to develop into a megacity by national government, a random planted seeds mechanism based on local maximums of transition potential is proposed to identify the potential development points in the local areas (neighborhood) of Xiongan New Area. Therefore, a 3×3 slide window is employed to pick out some local maximums of the urban transition potential surfaces before simulation. The local maximums was picked up by using the following rules:

$$\text{Localmax} = \begin{cases} \text{Pswin}_{(i=0,j=0)}^{n=3} > \max(\text{Pswin}_{(i\neq 0, j\neq 0)}^{n=3}) & -1 \leq i,j \leq 1 \\ \text{Pswin}_{(i=0,j=0)}^{n=3} > \epsilon \ \ or \ \ randn \times (1 - SearchWin_{urban}^{N'}) < \sigma \end{cases} \quad (8)$$

where Localmax means the local maximum of a slide window, if the transition potential of the central pixel of the slide window ($\text{Pswin}_{(i=0,j=0)}^{n=3}$) higher than the transition potential of other pixels ($\text{Pswin}_{(i\neq 0,j\neq 0)}^{n=3}$) in the 3×3 slide window, the central pixel is one of the local maximums of the transition potential surface. $\epsilon$ denotes a threshold within [0,1] defined by modeler to pick up the local maximums with higher urban transition potential. $randn$ is a random number between 0 and 1, and $SearchWin_{urban}^{N'}$ is an N'×N' search window, N' is an odd number. If $randn \times (1 - SearchWin_{urban}^{N'})$ less than a small unexpected threshold $\sigma$, the local maximum can also be planted in the geographic space. This condition allows the local maximums close to original urban land to develop into urban although the urban transition potential is low (lower than $\epsilon$), because the growing cities still need the support of original urban.

Then the random development seeds will be selected from these local maximums if the following conditions are fulfilled:

$$\text{Seeds} = \begin{cases} Amount_{seeds} < \frac{Demand_{urban} - Amound_{urban}}{APsize_{urban}} \\ SearchWin_{urban}^{N'} = 0 \quad N' \geq N, \; N' \bmod 2 \neq 0 \\ \text{Pswin}_{(i=0, j=0)}^{n=3} > r \quad 0 \leq r \leq 1 \end{cases} \quad (9)$$

where $Amount_{seeds}$ depicts the max amount of the random planted seeds, which is firstly estimated before simulation. $Demand_{urban}$ is the future urban amount specified by policy makers. $Amound_{urban}$ denotes the current urban amount. $APsize_{urban}$ represents the average size of the new urban patches, which is a input parameter of this model. The amount of the new planted seeds will increase until it reaches the max amount of the random planted seeds. Then the previously mentioned N'×N' search window is employed at the potential pixels of development seeds. If there are on any urban pixels in the search window, the next condition continue executes and a new seeds is possible to be planted. This condition is used to prevent the random planted seeds from getting too close to each other. Finally, given that the local maximums can also be found in where urban transition potential is very low. We need to give priority to the local maximums with high urban transition potential. Thus the third condition is implemented——when the transition potential of a local maximum ($\text{Pswin}_{(i=0, j=0)}^{3}$) is greater than a random value $r$ within (0, 1), a seed is planted on the cell. As a result of the random planted seeds, the transition potential is multiplied by a random value that replaces the neighborhood effect ($\Omega_{i,k}^t$ would likely be 0 in this condition) in formula (6) and obtains a total probability that are greater than 0. Then, the formula (6) is converted to following form:

$$TP_{i,k}^t = \begin{cases} P_{i,k} \times randn \times inertia_k^t \times con_{c \to k} & \text{if a Seed is planted} \\ P_{i,k} \times \Omega_{i,k}^t \times inertia_k^t \times con_{c \to k} & \text{others} \end{cases} \quad (10)$$

$randn$ is a random value between [0,1] which allows new urban to occur in where the neighborhood effect is zero. Figure 1 depicts a flow chart of the C-FLUS model.

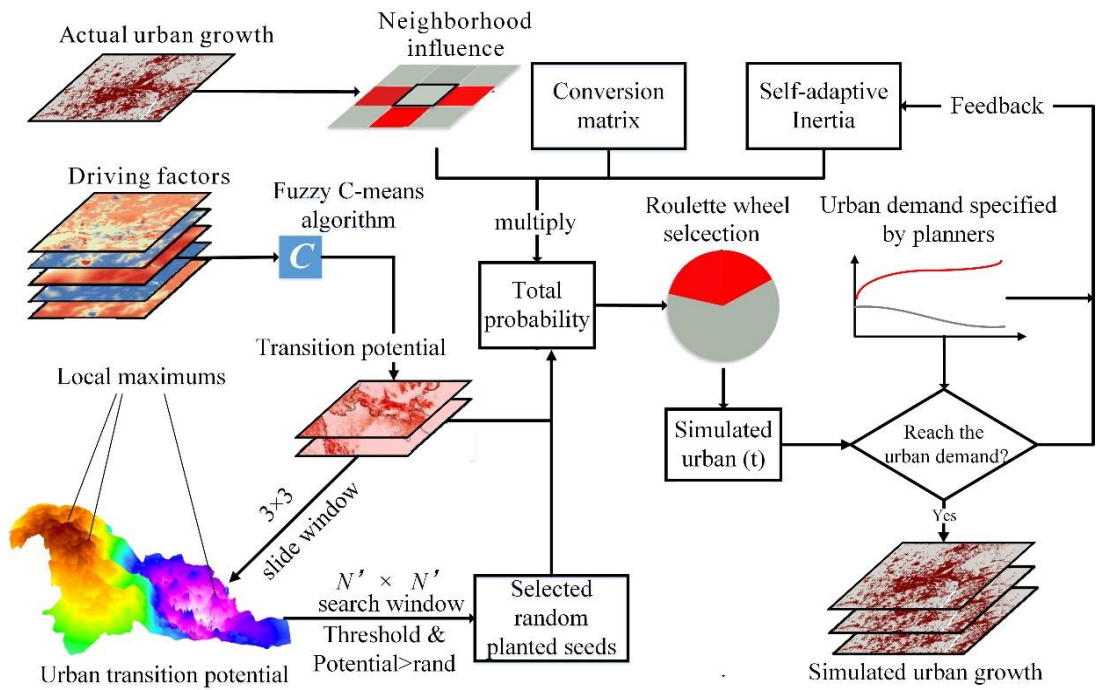

Figure 1. The framework of the proposed C-FLUS model for Xiongan New Area.

## 3. Study area and datasets

### 3.1. Study area

The Xiongan New Area is located in Hebei province in North China Plain, about 110

km southwest away from Beijing, the capital city of China. The study area of this research spans the 3 main counties: Xiongxian, Rongcheng and Anxin, which has a total area of 1572.36 km$^2$ and a residential population of approximately 1.047 million in June 2017 (Figure 2). The terrain in Xiongan New Area is generally flat, but relatively higher in the northwest and lower in the southeast. The largest lake in Hebei province, the Baiyang Lake, located in the southeast of the Xiongan New Area, which is a famous tourist attractions in China. Although the current urbanization level in this region are relatively low, Xiongan New Area is characteristic by its advantages of location for developing into a big city in several aspects includes deep soil layers, low vegetation coverage, convenient traffic, excellent ecological environment, high environmental carrying capacity and ample space for development (Jiang et al., 2017).

According to the government news, after the establishment of the Xiongan New Area, Chinese government planned to complete 6 railways, 4 high-speed railway stations and 1 airport within three years (from 2017 to 2020) inside or outside the Xiongan New Area for accelerating the construction of the new area. Although the official urban plan has not been determined, the development of Xiongan New Area is starting and more and more reliable messages are spread out by government sector, such as the siting of the planning high-speed railway stations and the designing of the high-speed railway line. At the same time, the land acquisition for temporary office block has already carried out in Rongcheng in September 2017. The temporary office block is probably the administrative center of Xiongan New Area in the future.

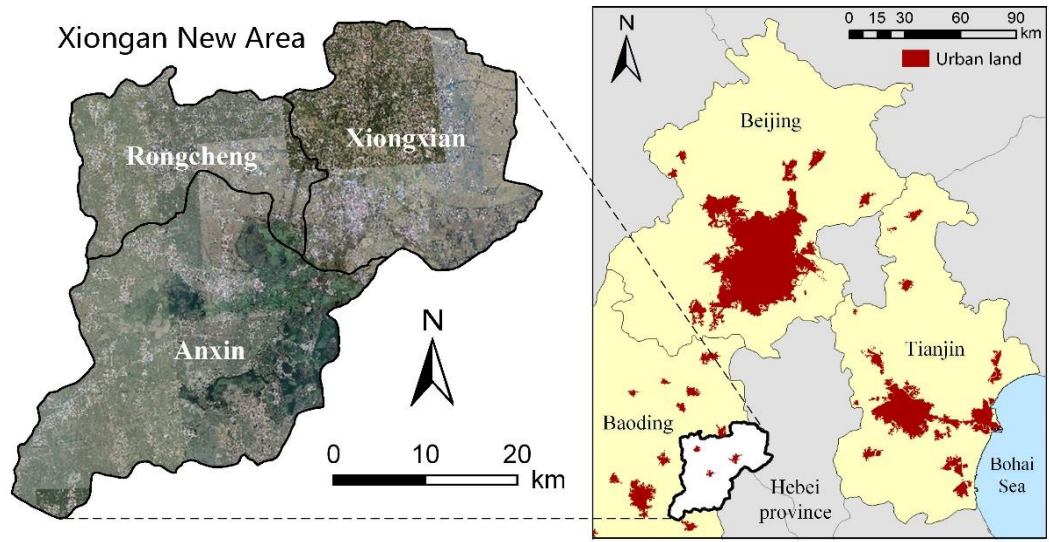

Figure 2. Spatial location of the Xiongan New Area.

**3.2. Data processing**

Together with these information, this research collected a series of driving factors that is closely related to the urban development and can reflect the present infrastructure construction in Xiongan New Area, including sites for development and the proximity to traffic network/hubs (e.g., highways, major roads, high speed railway stations and ports as well as proximity to governments) and socioeconomic factors such as proximity to vocational schools, villages and populated areas and etc. The traditional nature factors includes DEM, slope and natural scenic spots are considered in this study. Another factor that cannot be neglect is that the proximity to Beijing, the attraction of Beijing will not doubt guide the direction of urban growth in Xiongan. Besides, the distance to big cities around Xiongan, for example, Tianjin, and Baoding, are also take into consideration in mining the transition potential for urban land.

The urban land-use data were derived from visual interpretation of Landsat images in the years of 2013 with a 30-m resolution. We only considered two type of land use (urban and non-urban) in this study. The rural construction land is not regarded as urban land because the urban densities are two low in these region. The water area and a small amount of forest land are regarded as transition constraint and not allowed to convert to urban land. All the driving factors, such as the proximity to road, are calculated based on the 30-m resolution. Other driving factors with low resolution, such as Population (250m) and GDP (1000m), are resized to match the 30-m resolution with a resampling method. There are a total of 21 driving factors are considered, includes 19 current driving factors and 2 planning driving factors. The data in this study are listed in Table 1.

Table 1. List of data in this study and data sources

| Category | Data | Data resource | Year |
| --- | --- | --- | --- |
| Land use | Land use data | CAS (http://www.resdc.cn) | 2013 |
| Socioeconomic data | Population | http://data.jrc.ec.europa.eu/dataset/jrc-ghsl-ghs_pop_gpw4_globe_r2015a | 2015 |
| | GDP | http://www.geodoi.ac.cn/WebEn/Default.aspx | 2010 |
| | Populated area | Baidu Map API (http://apistore.baidu.com/) | 2017 |
| | Government | | |
| | Village | | |
| | Temporary office block | | |
| Infrastructure | Middle school | | |
| | Vocational school | | |
| | Primary school | | |
| | kindergarten | | |
| Location | Proximity to Beijing | | |
| | Proximity to around cities | | |

| Natural conditions | Scenic point | |
| --- | --- | --- |
| | DEM | GDEMDEM (http://www.gscloud.cn/) |
| | Slope | Calculated from DEM |
| Transportation | Waterway | Open Street Map (http://www.openstreetmap.org/) |
| | High-speed railway station | |
| | Railway and high-speed railway | |
| | Urban road network | |
| Planning data | Planning high-speed railway station | A Collection of Government Information |
| | Planning high-speed railway line | |

## 4. Model implementation and results

The applicability of the proposed C-FLUS model was first tested by discussing the urban transition potential exported by the FCM algorithm. Then the spatial simulation process of C-FLUS model is implemented then generate the simulation results in the mid-term urban amount (200 square km) decided by the Chinese government. The FCM algorithm divides the features listed in Table into two types: the potential urban and non-urban. The additional fuzzy exponent (m) is set as a value of 2.0. The termination threshold ($\varepsilon$) of FCM algorithm is set at $2.5 \times 10^{-15}$. The Euclidean distance is used to measure the distance between data points and cluster centers. In the simulation module, we used the $3 \times 3$ Moore neighborhood for the simulation based on the transition potential generated by different driving factors by using FCM algorithm. We compare the urban transition potential calculated by considering the planning high-speed railway systems with the transition potential without considering planning components in the following section.

**4.1. Urban development potential analysis**

Figure 3 depicts the urban development potential under the influences of planning high speed-railway and high speed railway stations (left) and the urban development potential without the influences of planning factors (right). A common feature of these two clustering results is that the highest development potential is emerge in the north areas of Xiongan New Area. It also reveals that the future urban land is least likely to appear in the southern region near the Baiyang Lake. However, at southwest corner near Baoding, the city closest to the Xiongan New Area, the urban development potential is relatively high but still much less than the northern development potential.

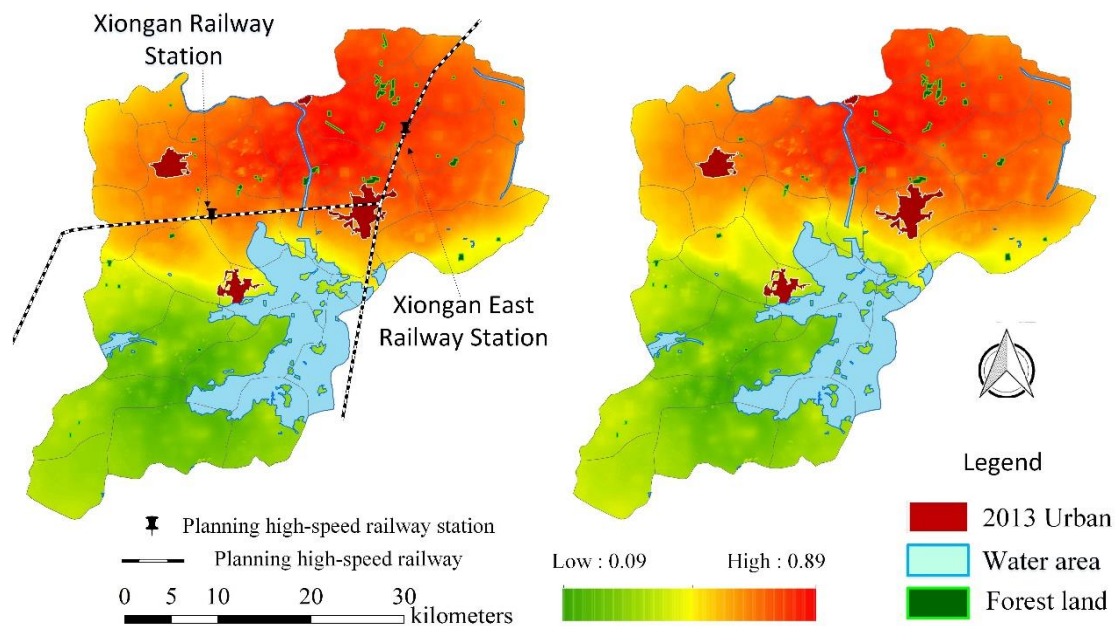

Figure 3. Urban transition potential in Xiongan New Area: transition potential under the influence of planning high-speed railway and high-speed railway stations (left);

transition potential without the influence of planning components (right).

In general, the urban development potential in Xiongan is lower in the south and higher in the north. The difference is that, under the influences of planning high-speed railway systems, the urban development potential under ther Planning Scenario stretches south to the region near the northern Baiyang Lake. They give more development chance for the central Xiongan New Area.

We divide the urban development potential into six level (Figure 4). The highest level of transition potential (>0.8257) occurs in the Daying town and Zhugezhuang town (Figure 4 right). If considering the driving effect of high-speed railway and high-speed railway stations, the Liangmatai town and Pingwang town will also have the highest level of transition potential (Figure 4 left). The second level of transition potential (0.7243-0.8257) is distributed around the highest level region. Under the influence of planning high-speed railway and high-speed railway stations, the second level of transition potential stretches into the area near the 2013 urban in northern Xiongxian and eastern Rongcheng. However, the original urban blocks in Xiongan New Area are mostly distributing in the third (Rongcheng and Xiongxian, 0.5683-0.7243) or fourth (Anxin, 0.3949-0.5684) level of transition potential. It means that Xiongxian has the most development chance among the three counties of Xiongan New Area, and the overall development potential in Rongcheng is higher than which in Anxin. The results also demonstrates that there is more appropriate region for building urban in Xiongan New Area.

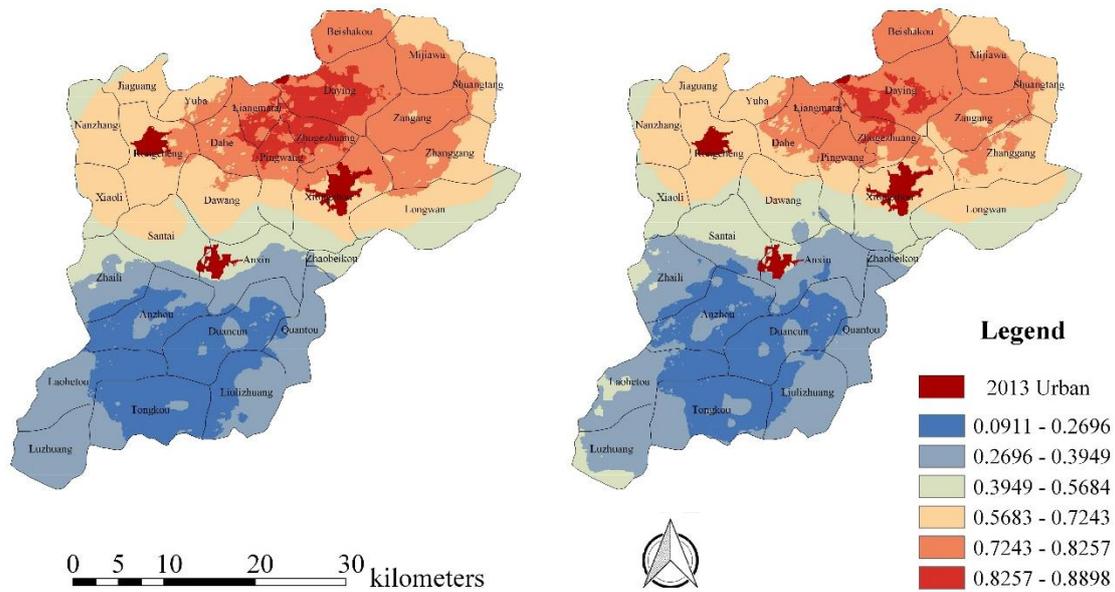

Figure 4. Classification for urban transition potential in Xiongan New Area: Classification for transition potential under the influence of planning high-speed railway and high-speed railway stations (left); Classification for transition potential without the influence of planning components (right).

## 4.2. Urban growth simulation

### 4.2.1. Model parameterization

Based on the transition potential exported by Fuzzy C-mean algorithm, the C-FLUS model is employed to simulate the future urban development in Xiongan New Area. The early stage development of the new urban area in Xiongan New Area is about 100 square km. It is expected that the Xiongan New Area will expand to 200 square km in the mid-term. Thus, the 100 and 200 square km is regarded as the future amounts of our simulation. However, the long-term planning control area of 2000 square km is too big for our start area (a total area of 1572.36 km$^2$). Therefore, the mid-long-term urban amount in this study is defined as twice the amount of mid-term urban area (400 square km), and the long-term urban amount is set to twice as much as the mid-long-term urban area (800 square km). Water area and forest land is regarded as the conversion constrain for protecting the future urban environment, which is not allowed to convert to urban area. The simulating parameters of the C-FLUS model is listed in Table 2:

Table 2. Simulating parameters of the C-FLUS model

| Input Parameters | Early stage | Mid-term | Mid-long-term | Long-term | Symbols |
|---|---|---|---|---|---|
| Urban amount (square km) | 100 | 200 | 400 | 800 | $Demand_{urban}$ |
| Slide window (pixel) | 3 (fixed) | 3 (fixed) | 3 (fixed) | 3 (fixed) | - |
| Neighborhood (pixel) | 3 | 3 | 3 | 3 | $N$ |
| Size of search window (pixel) | 21 | 21 | 21 | 21 | $N'$ |
| Average size of the new urban patches (pixel) | 100 | 100 | 100 | 100 | $APSize_{urban}$ |
| Threshold of local maximums (Scenario 1/2) | 0.84/0.82 | 0.77/0.80 | 0.75/0.78 | 0.50/0.53 | $\epsilon$ |
| Unexpected threshold | 0.01 | 0.01 | 0.01 | 0.01 | $\sigma$ |

We generate two urban expansion scenarios using the C-FLUS model (the Planning Scenario and the Baseline Scenario). Both scenarios have the same urban land area and simulation parameters in the mid-term and long-term period. Most of the simulating parameters in the three periods are the same except the local maximums threshold ($\epsilon$). Because a larger amount of future urban needs a lower threshold to allow urban growth in C-FLUS model. The urban transition potential under the Planning Scenario is higher than the urban transition potential of Baseline Scenario. Thus the thresholds of local maximums under the Planning Scenario (Scenario1: Early stage: 0.84, Mid-term: 0.80, Mid-long-term: 0.75, Long-term: 0.57) are higher than those in

Baseline Scenario (Scenario 2: Early stage: 0.82, Mid-term: 0.77, Mid-long -term: 0.72, Long-term: 0.54).

### 4.2.2. Simulation results in the Xiongan New Area

We used the observed actual 2013 urban land-use map as the start map of the simulation. Together with the corresponding simulation parameters under the specified simulation periods and scenarios, the future urban development in Xiongan New Area is projected by using the proposed C-FLUS model. Figure 5 illustrates the actual urban in 2013 and the simulation result under different scenarios in different simulation periods.

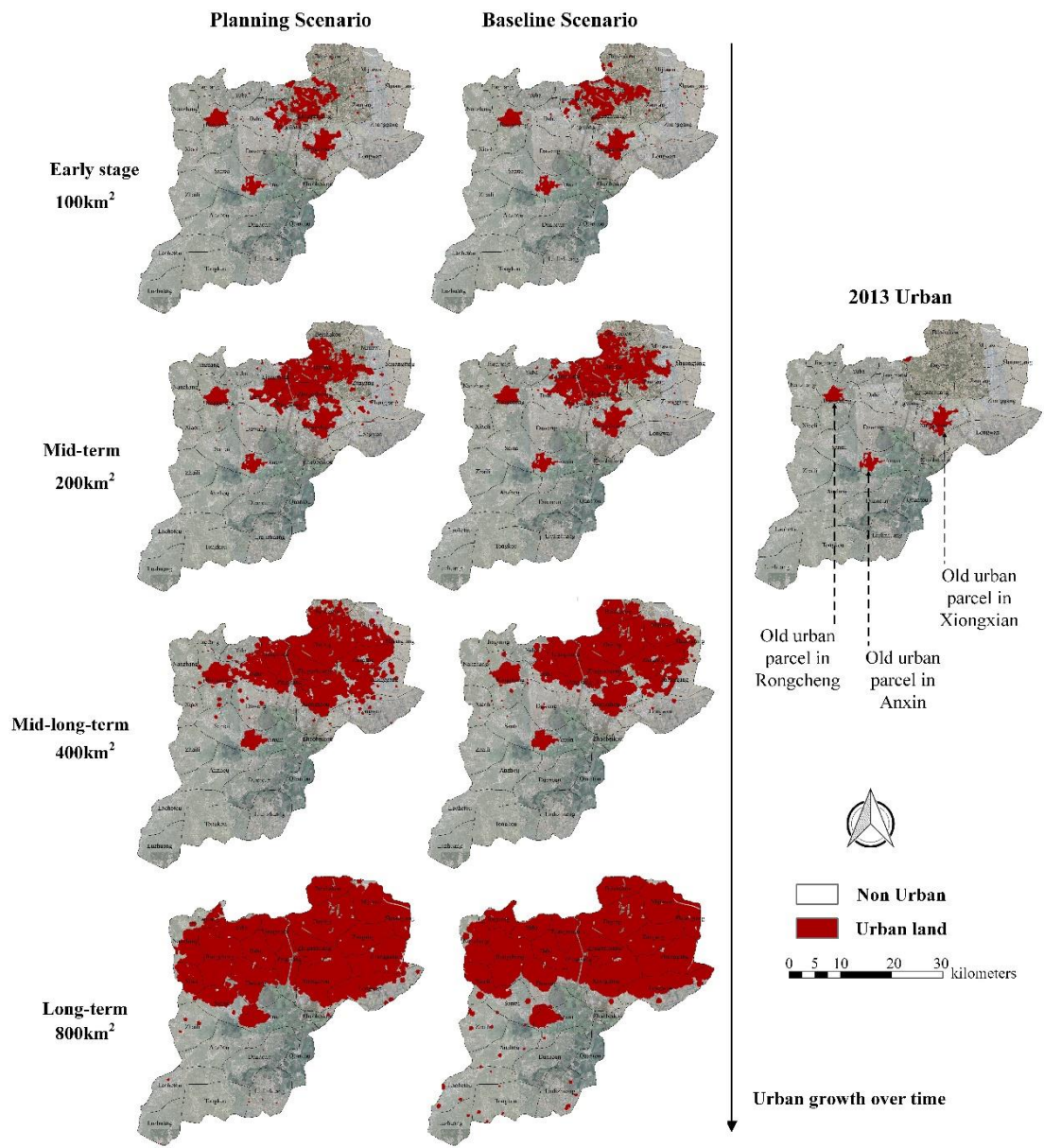

Figure 5. Simulated urban pattern in Xiongan New Area in different periods.

According to the simulation results in the early stage, the growing urban will firstly occur at the border of Xiongxian and Rongcheng. The new growth patches in the north will experience a rapid development and gather to form the predominant urban area. The towns such as Daying, Zhugezhuang, Liangmatai and northern part of Pingwang may be the first towns to developed new urban. In the mid-term simulation period, the urban development under the Planning Scenario depicts a stronger southward extension than the urban growth in Baseline Scenario. The new cities start to connect with the old urban parcel in Xiongxian, and more small urban patches around the old urban parcels in Xiongxian and Rongcheng are emerged in Xiongxian under the Planning Scenario. Instead, the urban development under the Baseline Scenario is characteristic by its extension to northeast corner of Xiongan New Area.

When the Xiongan New Area develops from mid-term period through to mid-long-term period, the urban land of the two scenario shows a relatively similar pattern. The small new urban patches emerge around Xiongxian and Rongcheng in the mid-term period have begun to take shape or develop into the big urban parcels until the mid-long-term period, and new growth patches are also be developed. In addition, the predominant urban land continue to develop south and tend to merge with the old urban blocks in Xiongxian. The northern new urban land is totally connecting with the old urban parcel in Xiongxian under the Planning Scenario. Meanwhile, westward and eastward development under the both scenarios is also very significant. The new urban land develops from east to west and starts to link with the western old urban parcel in Rongcheng under the Planning Scenario, while the urban land under the Baseline Scenario remains separate. This phenomenon demonstrates that the planning high-speed railway and high-speed railway stations will speed up the westward development of urban land in Xiongan New Area.

The long-term period simulation results show that when the urban amount reach 800km$^2$, the whole northern Xiongan New Area will be covered by urban area under the two scenarios. The further southward extension of the urban land is suppressed by the Baiyang Lake and the attraction of Beijing and Tianjin from the north and the east. Although Baoding is the nearest city to Xiongan New Area which located in south east of study region, its city attraction is too weak to guide the direction of urban growth in Xiongan New Area. In addition, the driving effect of the Xiongan railway station contributes to the connection between the northern predominant urban and the central old urban parcel in Anxin under the Planning Scenario. Instead, under the Baseline Scenario that without the influence of high-speed railway and high-speed railway stations , the old urban parcel in Anxin will remain isolated. However, the Baseline Scenario allows more disperse and small urban patches to be developed in long-term period. They will mainly distribute in southern Xiongan New Area around the Baiyang Lake.

The above results and analysis has shown that our proposed method can be employed to identify the potential development area in Xiongan New Area, and further predict the future urban form of Xiongan New Area. By considering current infrastructure and a little public planning information, the proposed method successfully project urban land use dynamics in Xingan New Area and obtains acceptable results in early stage, mid-term, mid-long-term and long-term simulation period. It also reveals that the planning high-speed railway and high-speed railway stations will speed up the connection between new urban land and old urban parcel in Xiongan New Area.

### 4.2.3. Analysis of candidate start zones

We compare our simulation results with three potential choices of start zones of Xiongan New Area proposed by Jiang et al. (2017). Three candidate start zones (the west candidate, the central candidate and the east candidate) for establishing the start-up new urban are analyzed in this research (Jiang et al., 2017). Similar with the point of our study that the urban development potential in Xiongan is lower in the south and higher in the north (Figure 3, 4), the three start zones are all located in the northern Xiongan New Area. The west candidate is located in the border of Rongcheng and Anxin, the central candidate is at the border of Xiongxian and Roncheng, and the east candidate is located in the northern part of Xiongxian.

In addition, Jiang et al. (2017) reported that the central candidate is the most recommended start zone for building new urban area in Xiongan New Area for its attention to overall development of the three counties and less demolition area. The second recommended start zone is the east candidate, which has the advantage of protecting high quality arable land. Our simulation results are in accordance with these conclusion. Figure 6 shows that, the simulation results of the C-FLUS model in the early stage is located in the region between the east candidate and the central candidate and cover a part of them respectively. Among the two scenarios, the Planning Scenario generates more urban land in the range of central candidate while the Baseline Scenario generates more urban land in the range of east candidate. The above analysis demonstrates that the simulation results of the proposed model is reasonable and available, which can provide more detailed and spatially explicit information for decision makers.

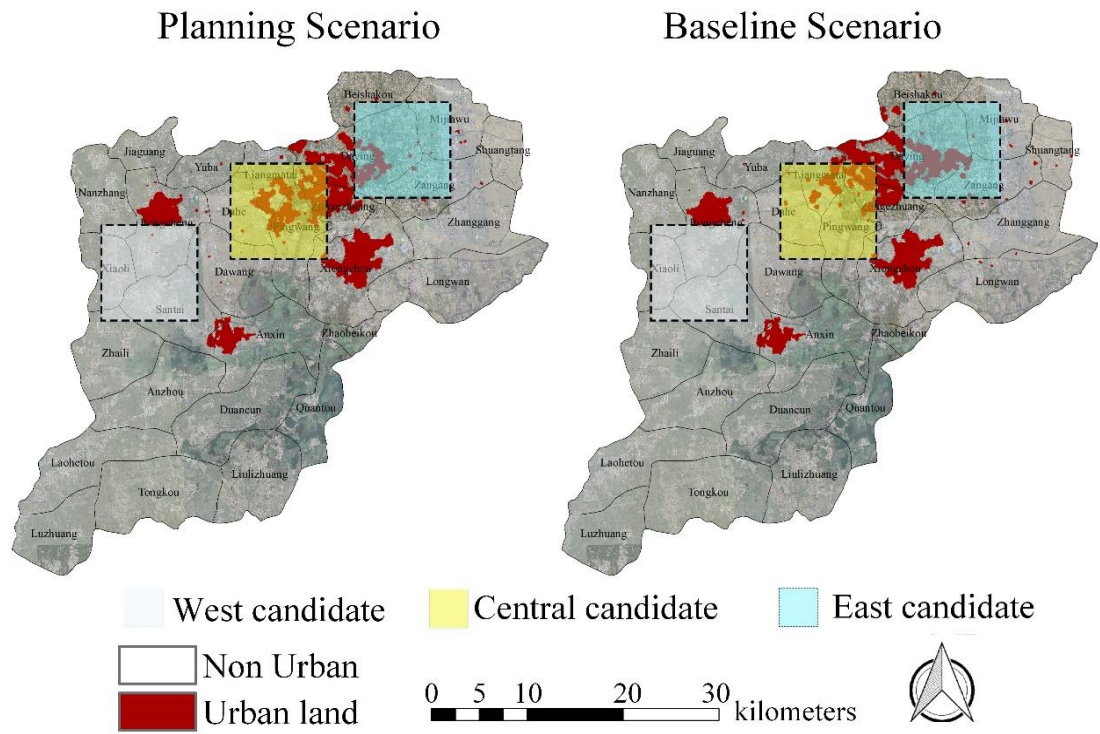

Figure 6. Compare the simulated urban pattern in the early stage with the candidate start zones proposed by Jiang et al. (2017)

## 5. Conclusion and discussion

In this study, we established a framework which is able to help national decision makers to discover the potential development area and predict probable urban pattern in Xiongan New Area. First, a soft unsupervised classification method (Fuzzy C-mean algorithm) was used to detect the urban development potential, in which various driving factors and planning policies are considered. In the next step, a C-FLUS model is proposed to simulate the future urban dynamics in Xiongan New Area by integrating a

CA-based simulation model (FLUS) with a random planted seeds mechanism based on local maximums of urban development potential. Moreover, the results of C-FLUS model effectively revealed complex urban development process in several simulation periods. The proposed model are validated through comparing the simulation results with the potential start zones proposed by previous study.

We find that the northern Xiongan New Area have a higher urban development potential than the south. The strong attraction of Beijing from the north and the constraint of the Baiyang Lake are important reasons that result in such distribution of development potential in Xiongan New Area. As a result, the new, high-density urban land will most likely to emerge in the town between northern old urban parcel in Xiongxian, most likely in Liangmatai, Zhugezhuang and Daying. The new urban land will expand in every direction with a much faster development speed than the old urban parcels. The new urban land will firstly merge the old urban parcel in Xiongxian, in north-eastern Xiongan New Area in the early stage and the mid-term period. Then from mid-term period to mid-long term period, the new urban land will westward develop and connect with the north-western old urban parcel in Rongcheng. Finally, the new urban land will occupy almost the whole northern Xiongan New Area and link with the southern old urban parcel in Anxin.

Two development scenarios are generated in this study. Compare to the Baseline Scenario that only considers current driving forces and infrastructure, the Planning Scenarios under the impact of planning high-speed railway and high-speed railway stations will promote the southward and westward development of new urban land in Xiongan New Area. It pays more attention to the North-central development of Xiongan New Area and contributes to the connection between new urban land and the old urban parcels. The planning high-speed railway systems will be more beneficial to meet the overall development of Xiongan New Area.

In general, this study explores a way to simulate urban growth in the place that almost have no high density urban and development basis. The main aim of this research was 1) to determine where to create the new urban, and 2) to predict how the future urban will evolve in Xiongan New Area based on the given planning information, current development basis and present infrastructure. It is expected that the proposed model can provide a better understanding of the urban dynamics and assist urban planners to explore future development alternatives for the area that mainly driven by the national policy. This method can not only employed in simulating the region that have any development basis, but also can be used to determine key zone for development for developed or developing cities. In our future study, the integration of the proposed method with more planning information will be conducted to satisfy the applications of urban studies in different scales in urban planning and city management.

**References:**

Bren D Amour, C., Reitsma, F., Baiocchi, G., Barthel, S., Güneralp, B., Erb, K., Haberl, H., Creutzig, F., Seto, K. C., 2017, Future urban land expansion and implications for global croplands, *Proceedings of the National Academy of Sciences* **114**(34)**:**8939-8944.

Chen, Y., Li, X., Liu, X., Ai, B., 2014, Modeling urban land-use dynamics in a fast developing city using the modified logistic cellular automaton with a patch-based simulation strategy, *International Journal of Geographical Information Science* **28**(2)**:**234-255.

Chen, Y., Li, X., Liu, X., Ai, B., Li, S., 2016, Capturing the varying effects of driving forces over time for the simulation of urban growth by using survival analysis and cellular automata, *Landscape & Urban Planning* **152:**59-71.

Chen, Y., Li, X., Wang, S., Liu, X., Ai, B., 2013, Simulating Urban Form and Energy Consumption in the Pearl River Delta Under Different Development Strategies, **103**(6)**:**1567 - 1585.

Clarke, K. C., 2014, Why simulate cities? *GeoJournal* **79**(2)**:**129-136.

Clarke, K. C., Gaydos, L. J., 1998, Loose-coupling a cellular automaton model and GIS: long-term urban growth prediction for San Francisco and Washington/Baltimore., *International Journal of Geographical Informationence* **12**(7)**:**699-714.

Deal, B., Schunk, D., 2004, Spatial dynamic modeling and urban land use transformation: a simulation approach to assessing the costs of urban sprawl,

*Ecological Economics* **51**(1-2)**:**79-95.

Fragkias, M., Seto, K. C., 2009, Evolving rank-size distributions of intra-metropolitan urban clusters in South China, *Computers, Environment and Urban Systems* **33**(3)**:**189-199.

Guy, E., Roger, W., Inge, U., 1997, Integrating Constrained Cellular Automata Models, GIS and Decision Support Tools for Urban Planning and Policy Making, pp. 125-155.

Jiang, L., Peiyi, L. V., Feng, Z., Liu, Y., 2017, Land use patterns of the Xiongan New Area and comparison among potential choices of start zone, *Resources Science*.

Jokar Arsanjani, J., Fibæk, C. S., Vaz, E., 2018, Development of a cellular automata model using open source technologies for monitoring urbanisation in the global south: The case of Maputo, Mozambique, *Habitat International* **71:**38-48.

Kamusoko, C., Gamba, J., 2015a, Simulating Urban Growth Using a Random Forest-Cellular Automata (RF-CA) Model, *ISPRS International Journal of Geo-Information* **4**(2)**:**447-470.

Kamusoko, C., Gamba, J., 2015b, Simulating Urban Growth Using a Random Forest-Cellular Automata (RF-CA) Model, *ISPRS International Journal of Geo-Information* **4**(2)**:**447-470.

Ke, X., Zheng, W., Zhou, T., Liu, X., 2017, A CA-based land system change model: LANDSCAPE, *International Journal of Geographical Information Science* **31**(9)**:**1798-1817.

Li, X., Chen, G., Liu, X., Liang, X., Wang, S., Chen, Y., Pei, F., Xu, X., 2017, A New Global Land-Use and Land-Cover Change Product at a 1-km Resolution for 2010 to 2100 Based on Human–Environment Interactions, *Annals of the American Association of Geographers* **107**(5)**:**1040-1059.

Li, X., Li, L., Lu, H., Chen, D., Liang, Z., 2003, Inhomogeneity correction for magnetic resonance images with fuzzy C-mean algorithm, SPIE, pp. 995 - 1005.

Li, X., Liu, X., Yu, L., 2014, A systematic sensitivity analysis of constrained cellular automata model for urban growth simulation based on different transition rules, **28**(7)**:**1317 - 1335.

Li, X., Yeh, A. G., 2002, Neural-network-based cellular automata for simulating multiple land use changes using GIS, *International Journal of Geographical Information Science* **16**(4)**:**323-343.

Li, X., Yeh, G. O., 2001, Calibration of Cellular Automata by Using Neural Networks for the Simulation of Complex Urban Systems, *Environment & Planning A* **33**(8)**:**1445-1462.

Liu, X., 2006, An extended cellular automaton using case–based reasoning for simulating urban development in a large complex region, *International Journal of Geographical Information Science* **20**(10)**:**1109-1136.

Liu, X., Hu, G., Ai, B., Li, X., Tian, G., Chen, Y., Li, S., 2017, Simulating urban dynamics in China using a gradient cellular automata model based on S-shaped curve evolution characteristics, *International Journal of Geographical*

*Information Science* **0**(0)**:**1-29.

Liu, X., Liang, X., Li, X., Xu, X., 2017, A future land use simulation model (FLUS) for simulating multiple land use scenarios by coupling human and natural effects, *Landscape and Urban Planning* **168:**94-116.

Rahimi, S., Zargham, M., Thakre, A., Chhillar, D., 2004, A parallel Fuzzy C-Mean algorithm for image segmentation, pp. 234-237 Vol.1.

Selvakumar, J., Lakshmi, A., Arivoli, T., 2012, Brain tumor segmentation and its area calculation in brain MR images using K-mean clustering and Fuzzy C-mean algorithm, pp. 186-190.

Sohl, T., Sayler, K., Drummond, M., Loveland, T., 2007, The FORE-SCE model: a practical approach for projecting land cover change using scenario-based modeling, *Journal of Land Use Science* **2**(2)**:**103-126.

Verburg, P. H., Schot, P. P., Dijst, M. J., Veldkamp, A., 2004, Land use change modelling: current practice and research priorities, *Geojournal* **61**(4)**:**309-324.

Verburg, P. H., Soepboer, W., Veldkamp, A., Limpiada, R., Espaldon, V., Mastura, S. S., 2002, Modeling the spatial dynamics of regional land use: the CLUE-S model., *Environmental Management* **30**(3)**:**391-405.

Yang, Q., Li, X., Shi, X., 2006, Cellular automata for simulating land use changes based on support vector machines, *Journal of Remote Sensing* **34**(6)**:**592-602.

Yao, Y., Liu, X., Li, X., Liu, P., Hong, Y., Zhang, Y., Mai, K., 2017, Simulating urban land-use changes at a large scale by integrating dynamic land parcel subdivision and vector-based cellular automata, *International Journal of Geographical Information Science* (2).

Yi-Qing, W. U., Liu, T. E., 2017, Consideration about the Xiongan New Area to Reasonably Undertake Beijing Non-capital Function, *West Forum*.